\documentstyle[twocolumn,aps]{revtex}

\begin{document}
\draft
\title{Longitudinal spin transport in diluted magnetic semiconductor superlattices:
the effect of the giant Zeeman splitting}
\author{Kai Chang\cite{kai}, J. B. Xia}
\address{National Laboratory for Superlattices and Microstructures,\\
Institute of Semiconductors,\\
Chinese Academy of Sciences,\\
P. O. Box 912, 100083, Beijing, China}
\author{F. M. Peeters\cite{francois}}
\address{Departement Natuurkunde, Universiteit Antwerpen (UIA), Universiteitsplein 1,%
\\
B-2610 Antwerpen, Belgium}
\maketitle

\begin{abstract}
Longitudinal\ spin transport in diluted magnetic semiconductor
superlattices\ is investigated theoretically. The longitudinal
magnetoconductivity (MC) in such systems exhibits an oscillating behavior as
function of an external magnetic field. In the weak magnetic field region
the giant Zeeman splitting plays a dominant role which leads to a large
negative magnetoconductivity. In the strong magnetic field region the MC
exhibits deep\ dips with increasing magnetic field. The oscillating behavior
is attributed to the interplay between the discrete Landau levels and the
Fermi surface. The decrease of the MC at low magnetic field is caused by the 
$s-d$ exchange interaction between the electron in the conduction band and
the magnetic ions.
\end{abstract}


\section{\bf INTRODUCTION}

The most striking phenomena in semiconductor quantum structures is the
tremendous change of the optical and transport properties induced by the
quantum confinement effect. Use of diluted magnetic semiconductors (DMS) in
such systems provides us with a new degree of freedom to engineer the
optical and transport properties by applying an external magnetic field\cite
{review}. An external magnetic field magnetizes the magnetic ions, which
gives rise to the exchange field acting on the electron spin. This
spin-dependent energy shift is comparable to the band-offset in a DMS
superlattice, therefore influencing significantly the optical property of
DMS.\ The optical properties of the DMS systems have been studied
extensively in the past few years. Time-resolved photoluminescence of a
dilute magnetic semiconductor superlattice has shown the feasibility of the
spin-alignment mechanism\cite{Dai1,Dai,Lunn}. The strong $s-d$ exchange
interaction between the electron spin in the conduction band and the
localized magnetic ions gives rise to unique magneto-optical properties\cite
{Kossut,Furdyna,von}. Giant Zeeman splitting, excitonic magnetic polaron,
Faraday rotation and optically induced magnetization\ are well known
examples. Recent experiments demonstrated that spin polarized transport in
diluted magnetic semiconductors and spin coherence can be maintained over
large distances ($\geq 100\mu m$) and for long times ($10^{-9}-10^{-8}s$)\
in metals and semiconductors and showed that the spin of the electron offers
unique possibilities for quantum computation and information transmission 
\cite{SPT,Kikkawa,Fiederling,Ohno}.

One of the fascinating effects of magnetic fields on the electron transport
properties in bulk materials is the well-known Shubnikov-de Haas (SdH)
effect, i.e., magnetoconductivity (MC) or magnetoresistance (MR) of the
system is independent of the magnetic field strength at very low magnetic
field, and exhibits an oscillating dependence with the magnetic field
strength\ at higher magnetic fields.\cite{Lend} This phenomena arises from
the interplay between the quantized Landau levels and the Fermi energy. In
semiconductor superlattices the SdH effect displays a rich diversity of
prominent phenomena since the electron motion along the magnetic field is
quite different from that in bulk materials. The conductivity of a system is
determined by the number of different states near the Fermi energy, the
group velocity associated with them, and the coupling of these states to
each other by scattering mechanisms. Polyanovskii\cite{Poly} presented a
theory to describe longitudinal magnetotransport in semiconductor
superlattices using the semiclassical approach. A single-band tight-binding
model was used to describe the superlattice at very high and very low
temperature. Datars and Sipe\cite{Datar} extended the theory to the multiple
miniband case, and found multiple miniband oscillations in the regime where
the second miniband is populated.

In this paper we focus on the effect of the giant Zeeman splitting on the
longitudinal magnetoconductivity in DMS superlattices and take as an example
the ZnSe/Zn$_{0.96}$Mn$_{0.04}$Se superlattices (see Fig. 1). In a DMS
system, the giant Zeeman splitting induced by the s-d exchange interaction
is comparable to the band offset, and therefore can change significantly the
energy spectrum of the minibands and the group velocity, which influences
the magnetoconductivity. Here we extend the treatment of the SdH effect from
ordinary semiconductor superlattices\ to diluted magnetic semiconductor
superlattices. We find a spin-dependent conductivity when an external
magnetic field is applied along the growth direction. The
magnetoconductivity decreases significantly with increasing magnetic field
at low magnetic fields, and exhibits an oscillating behavior in strong
magnetic fields. A strong spin polarized current is found with increasing
magnetic field. The underlying physics arises from the {\it s-d} exchange
interaction between the intinerant electron and the localized magnetic
impurity which lifts the degeneracy of the spin-up and spin-down electron
band states. Here we neglect the effect of the spin-flip process\cite
{Bastard} and focus only on the effect of the giant Zeeman splitting caused
by the {\it s-d} exchange interaction on the MC\ in a DMS superlattice for
electrical transport along the growth axis. The numerical results obtained
within this approximation were recently found to be in excellent agreement
with SdH measurement performed on Cd$_{1-x}$Mn$_x$Te/Cd$_y$Mg$_{1-y}$Te
quantum wells (Fig. 4 in Ref. \cite{ACC}).

The paper is organized as follows, the theoretical model is described in
Sec. II, and the numerical results and discussions are given in Sec. III.
Finally, the conclusion is presented in Sec. IV.

\section{\bf THEORETICAL MODEL}

We model a DMS superlattice as a periodic array of square potential wells
and non magnetic barriers and assume that the magnetic ions are distributed
homogeneously in the DMS layers (see Fig. 1). In ordinary semiconductor
superlattices, the Zeeman splitting is quite small due to the small
Land\'{e} $g$ factor. In a DMS superlattice, a small external magnetic field
gives rise to a giant Zeeman splitting of the conduction band states, and
results in striking differences between spin-up and spin-down\ electron
states of the system. This giant Zeeman splitting arises from the spins of
the injected electrons interacting with the $S=5/2$ spins of the localized $%
3d^5$ electrons of the Mn$^{2+}$ ions via the {\it s-d} exchange interaction%
\cite{Kossut}. Our theory is based on the assumption that the electron
motion in the DMS superlattice can be well described by the effective mass
approximation (EMA) which are confirmed by recent experiments\cite
{Rossin,SLee,ACC}. As shown in Fig. 1, the model Hamiltonian for electrons
in such a system is 
\begin{eqnarray}
H &=&p_x^2/2m_e^{*}+(p_y+eBx)^2/2m_e^{*}+p_z^2/2m_e^{*}  \nonumber \\
&&+V_{conf}(z)+J_{s-d}\sum_i{\bf s}({\bf r}){\bf \cdot S}({\bf R}_i)\delta (%
{\bf r}-{\bf R}_i).\ ,  \label{Hamil}
\end{eqnarray}
where $V_{conf}(z)=V_{conf}(z+L)$ is the periodic potential along the growth
direction, $L$ is the period of the DMS\ superlattice, and ${\bf S}$ is the
spin of the localized $3d^5$ electrons of the Mn ions with $\ S=5/2$ and $%
{\bf s}$ is the electron spin. We assume that the magnetic ions are
distributed homogeneously in the DMS layers. The extended nature of the
electronic wave function spanning a large number of magnetic ions and
lattice sites allows the use of the molecular-field approximation to replace
the magnetic-ion spin operator ${\bf S}_i$ with its thermal and spatial
average $<S_z>$, taken over all the ions. The mean spin $<S_z>$ denotes the
spatial as well as the thermal average of the spin component along the
magnetic field. This approach has been proven to be valid in previous
theoretical works on magneto-optical property of DMS material. Very
recently, this approximation was applied sucessfully to study the transport
property of DMS\ system.\cite{ACC,Egues} The exchange interaction given by
Eq. (\ref{Hamil}) can, in the molecular-field approximation, be written in
terms of a Zeeman-like Hamiltonian 
\begin{eqnarray}
H &=&p_x^2/2m_e^{*}+(p_y+eBx)^2/2m_e^{*}  \nonumber \\
&&+p_z^2/2m_e^{*}+V_{conf}(z)+J_{s-d}<S_z>s_z\ ,  \label{meanfield} \\
&=&p_x^2/2m_e^{*}+(p_y+eBx)^2/2m_e^{*}+p_z^2/2m_e^{*}+V_{eff}(z)\ , 
\nonumber
\end{eqnarray}
where$\ J_{s-d}=-N_0\alpha x_{eff}$ and $<S_z>=5/2B_J(Sg\mu _BB/k_B(T+T_0))$%
, $B_J(x)$ is the Brillouin function, $N_0$ is the number of cations per
unit volume, the phenomenological parameters $x_{eff}$ (reduced effective
concentration of Mn) and $T_0$ accounts for the reduced single-ion
contribution due to the antiferromagnetic Mn-Mn coupling, $k_B$ is the
Boltzmann constant and $s_e$ is the electron spin. The parameters used in
the calculation are taken from Ref. \cite{Dai}: $m_e^{*}=0.16m_0$, $%
V_{conf}=-3meV$ in the DMS material, $g=2$, $g_s=1.1$, $N_0\alpha =0.27eV$, $%
T_0=1.4K$. In Eq. (\ref{meanfield}) the exchange interaction $%
H_{s-d}=J_{s-d}<S_z>s_z$ only induces spin-conserving processes, and
consequently we neglected all spin-flip processes.

Using the usual boundary conditions\cite{Burt} for the electron wavefunction
at the well/barrier interface, the energy eigenvalue can be obtained by
solving the following equation 
\begin{eqnarray}
\cos k_zL &=&\cos k_N^sL_N\cos k_D^sL_D-\frac 12(\frac{m_Dk_N}{m_Nk_D}+\frac{%
m_Nk_D}{m_Wk_N})\ , \\
&&\times \sin k_N^sL_N\sin k_D^sL_D  \nonumber
\end{eqnarray}
where $k_N^s=\sqrt{2m_N(E-V_N^s)/\hbar ^2}$, $k_D^s=\sqrt{%
2m_D(E-V_D^s)/\hbar ^2}$, and $V_D^s=V_{conf}\pm 1/2N_0\alpha x_{eff}$ $%
<S_z> $ is the depth or the height of the DMS\ layer which depends on the
spin orientation($s=\pm $, $+$for the spin-up electron, $-$for the spin-down
electron). The period of the superlattice is $L=L_D+L_N$, where $L_D$ is the
width of the DMS layers and $L_N$ is the width of the nonmagnetic
semiconductor layers. $m_D$ and $m_N$ denote the effective mass of the
electron in the DMS layers and nonmagnetic semiconductor layers,
respectively. In this work the difference between the effective mass of the
electron in the DMS\ layers and the non DMS\ layers is neglected which is a
reasonable assumption because of the low Mn concentration ( $m_N=m_D=m_e^{*}$%
). Notice that the barrier height and the well depth can be tuned by varying
the external magnetic field.

In the presence of a magnetic field, the in-plane motion of the electron is
described by discrete Landau levels on which any effect of the collision
broadening is ignored.\ Therefore, the eigenvalue of the electron state in a
DMS\ superlattice under a perpendicular magnetic field is 
\begin{equation}
{\cal E}(\tilde{k})=(n+\frac 12)\hbar \omega _c+{\cal E}(k_z)\ ,
\label{energy}
\end{equation}
where $\tilde{k}=(n,k_z,s)$ is the complete set of quantum indices, $\omega
_c=eB/m_e^{*}$ is the cyclotron frequency, and $n$ denotes the label for the
Landau level. ${\cal E}(k_z)$ is the energy spectrum of the miniband in the
DMS\ superlattice.

The group velocity of the electron along the $z$-direction is

\begin{equation}
v_{s,z}=\frac 1\hbar \frac{\partial E_s(n,k)}{\partial k},\ s=\uparrow
,\downarrow .  \label{v_g}
\end{equation}

The ballistic current density $J^s$ is the sum of the contributions from
each Landau level with different spin 
\begin{equation}
J=-ne<v_z>=-\frac{eE}{2\pi l_B^2}\sum_{s,n}\int_{-\pi /L}^{\pi /L}\frac{dk_z%
}{2\pi }v_{s,z}(n,k)f_s(n,k),\ \ \ \ \   \label{current}
\end{equation}
where $1/2\pi l_B^2$ is the degeneracy of the Landau level for each spin, $%
l_B=\sqrt{\hbar /eB}$ is the magnetic length, $f_s(n,k)$ is the state
distribution function which can be determined from the semiclassical
Boltzmann equation 
\begin{equation}
\frac{\partial f_s}{\partial t}+{\bf v}\cdot \nabla f_s+\frac{\partial {\bf k%
}}{\partial t}\cdot \nabla _kf_s=(\frac{\partial f_0}{\partial t})_c.
\label{Boltz}
\end{equation}
If the distribution $f_s$ depends weakly on the position $z$ along the
growth direction, and is independent of the time, Eq. (\ref{Boltz}) becomes 
\begin{equation}
-e{\bf E}\cdot \nabla _kf_s=(\frac{\partial f_0}{\partial t})_c.
\label{coll}
\end{equation}
If we use the relaxation time approximation, the collision term on the right
side of Eq. (\ref{coll}) is equal to 
\begin{equation}
(\frac{\partial f_0}{\partial t})_c=-\frac{f(\tilde{k})-f_0(\tilde{k})}{\tau
(\tilde{k})}\ ,
\end{equation}
where $f_0({\cal E}(\tilde{k}))=1/[\exp (({\cal E}(\tilde{k})-E_F)/k_BT)+1]$
is the Fermi-Dirac distribution, $E$ is the electric field along the growth
direction and $\tau (\tilde{k})$ denotes the electron relaxation time in the
DMS\ superlattice.

The Fermi energy can be determined from the following equation 
\begin{equation}
n_e=\frac{e^2}{4\pi ^2l_B^2}\sum_{s,n}\int_{-\pi /L}^{\pi /L}f({\cal E(}%
n,k_z,s{\cal )})dk_z\ .  \label{fermi}
\end{equation}
where $n_e$ is the density of the electron in the DMS\ superlattice.

We restrict ourselves to the linear-response regime, assume weak electric
field, and ignore spin-flipping processes, therefore the distribution
function can be written in the form of $f=f_0+f_1=f_0-ev\tau E\partial
f_0/\partial {\cal E}$, here $f_0$ is the equilibrium distribution function
and $f_1$ is the linear term which is proportional to the electric field.
Because there is no electric current in the equilibrium Fermi-Dirac
distribution, the current density Eq. (\ref{current}) becomes

\begin{equation}
J=\frac{e^2E\tau }{4\pi ^2l_B^2}\sum_{s,n}\int_{-\pi /L}^{\pi /L}dk_z({\bf -}%
\frac{\partial f_0}{\partial {\cal E}})v_{s,z}^2.\ \ \ \ \ 
\end{equation}
From this formula, we can find that the current density is ascribes to the
contribution of the Landau level near the Fermi energy, and especially at
low-temperature ${\bf -}\frac{\partial f_0}{\partial {\cal E}}\approx \delta
({\cal E}-{\cal E}_F)$ for $k_BT\ll {\cal E}_F$. The conductivity $\sigma $
can be obtained as 
\begin{equation}
\sigma =\frac{e^2\tau }{4\pi ^2l_B^2}\sum_{s,n}\int_{-\pi /L}^{\pi /L}dk_z(%
{\bf -}\frac{\partial f_0}{\partial {\cal E}})v_{s,z}^2.  \label{conduc}
\end{equation}

The degree of spin polarization of the current density under weak electric
field can be defined as 
\begin{equation}
P=\frac{J^{\downarrow }-J^{\uparrow }}{J^{\downarrow }+J^{\uparrow }}\ ,
\label{Polarization}
\end{equation}
here $J^{\uparrow }(J^{\downarrow })$ is the component of spin-up
(spin-down) current density.

\section{\bf NUMERICAL RESULTS AND DISCUSSIONS}

Figures 2 (a) and 2 (b) show the energy spectrum of the lowest two miniband
spin-up (solid curves) and spin-down (dashed curves) electron states (see
Eq. (\ref{energy}), $n=0$) for a ZnSe/Zn$_{0.96}$Mn$_{0.04}$Se diluted
magnetic\ semiconductor superlattice for different magnetic fields. From
these figures we notice that the separation between the spin-up and
spin-down electron is enhanced significantly with increasing magnetic field.
Notice that the Fermi energy is located slightly above the bottom of the
second miniband at low magnetic fields (see Fig. 2 (a)). In strong magnetic
fields, the energy of the lowest spin-up miniband is even higher than that
of the spin-down second miniband, and only the lowest spin-down miniband is
occupied by electrons (see Fig. 2 (b)). This can be explained as follows, an
external magnetic field induces a magnetization of the magnetic ion Mn$^{2+}$
along the direction of the magnetic field in the DMS superlattice. From Eq. (%
\ref{Hamil}), the magnetic ions can influence the energy of the electron
state via the {\it s-d} exchange interaction, and leads to a giant spin
splitting which is comparable to the band offset between ZnSe and Zn$_{0.96}$%
Mn$_{0.04}$Se.

Figs. 2(c) and 2(d) depict how the bandwidths of the electrons for different
spin orientation vary with the magnetic field. For spin-down electrons the
bandwidth decreases and saturates with increasing magnetic field, but for
spin-up electrons, it exhibits a maximum and saturates when the magnetic
field increases. For spin-up electrons the wells become more and more
shallow and finally form barriers with increasing magnetic field. At this
point the band widths diverge and the miniband gaps disappear (see Fig.
2(c)). Therefore the band width for the spin-up electron exhibits a local
maximum. In Fig. 2(e) we plot the group velocity of the electrons with
different spin orientation as a function of the momentum $k_z$ for different
magnetic fields. At low fields the group velocity for the spin-up electron
is larger than that in the absense of magnetic field. The group velocity for
the spin-up and spin-down electrons decrease at strong magnetic fields.

Figure 3 shows the Landau level fan diagram for spin-down (thick curves) and
spin-up (thin curves) electrons in a DMS superlattice under a perpendicular
magnetic field. The solid (dashed) lines indicate the energy of the Landau
level at $k_z=0$ ($k_z=\pi /L_z$), i.e. the edge of the lowest miniband.
Notice that the magnetic field variation of the Landau levels in DMS is
quite different from that in a nonmagnetic semiconductor, which is linearly
proportional to the external magnetic field. In a DMS\ superlattice the
Landau levels exhibit minima with increasing magnetic field, and its
variation with magnetic field is quite different from that of the Landau
levels for the spin-up electron. For very small magnetic fields the {\it s-d}
exchange interaction increases the barrier height for the spin-up electrons
moving the Landau levels up in energy, while for the spin-down electrons the
wells deepen resulting in a decrease of the Landau level energy. This effect
saturates around $B\sim 4T$ when the magnetization of the Mn$^{2+}$ is
saturated beyond which we have the usual Landau level fan diagram for each
of the electron spin states. The spin-up and spin-down fan are shifted in
energy due to the fact that they move in a different superlattice potential.
The thickest solid curve in Fig.~3 denotes the Fermi energy vs the magnetic
field. Sharp drops take place at the points where the Fermi energy passes
through the bottom of the different Landau level bands. From this figure we
also learn that the electron state becomes spin-polarized since only the
lowest spin-down miniband is populated at sufficient large fields.

Figure 4 shows the conductivity $\sigma $ as a function of magnetic field
for various temperatures in a DMS superlattice. The inset shows the Fermi
energy vs magnetic field in such a system. An interesting property of the MC
is the variation of the low-field MC. The conductivity decreases and
oscillates with increasing magnetic field. At low magnetic field, the spin
splitting induced by the {\it s-d} exchange interaction is even much larger
than the separation of the Landau levels, the {\it s-d} exchange interaction
results in a variation of the miniband width, i.e. a variation of the
electron group velocity (see Eq. (\ref{v_g})). From Fig. 1(c) we find that
an increase of the magnetic field leads to a decrease of the bandwidth for
the spin-down and the spin-up electron and a local maximum for the spin-up
electron, i.e. a decrease of the group velocity for the spin-down and
spin-up electron and a local maximum for the spin-up electron (see Fig.
2(e)). Therefore, the MC exhibits a maximum for the spin-up electron and a
decrease for the spin-down electron in the low-field case (see Fig. 6). The
decrease of the low-field MC was found previously in disordered 2D system
which was attributed to quantum corrections caused by Anderson weak
localization \cite{PALee}. But the decrease of MC in a DMS\ superlattice
arises from the {\it s-d} exchange interaction between the intinerant
electron and the localized magnetic impurity which lifts the degeneracy of
the spin-up and spin-down electron band states.

The magnetization of the magnetic ions Mn$^{2+}$ saturates with increasing
magnetic field, therefore the strength of the exchange interaction (the last
term in Eq. (\ref{meanfield})) also saturates when the magnetic field
becomes strong enough (B%
\mbox{$>$}%
4T). The separation of the Landau levels increases linearly (see Fig. 3)
with increasing magnetic field. The Fermi surface passes through the band
bottom of the subsequent Landau levels with increasing magnetic field. The
Fermi energy (see the inset) decreases and shows a series of sharp drops at
strong magnetic field. The variation of MC in strong magnetic field is
ascribed to the contribution from Landau levels near the Fermi energy. When
a Landau level passes through the Fermi surface, the electron group velocity
of the states which contribute to conduction drops to zero (see Eq. (\ref
{conduc})) resulting in an oscillation of the MC. The conductivity exhibits
a sharp dip if there is only one Landau level near the Fermi energy. The
separation of the Landau levels is small at low field and these dips are
smeared out since there are many Landau levels located near the Fermi
surface. From this figure, we can also see that the dips will be less
pronounced when temperature increases since the latter leads to a smearing
of the Fermi surface.

Figure 5 depicts how the conductivity $\sigma $ varies with magnetic field
for different carrier density. The period of the conductivity oscillations
for lower density is larger than that for higher density, which can be
understood from the inset which shows the variation of the Fermi energy
versus magnetic field. When the Landau level passes through the Fermi
surface, a corresponding dip can be found in the conductivity. Since the
period of the oscillation of the Fermi energy for lower density is also
larger than that for higher density, the period of the MC oscillation for
lower density will be larger than for higher density.

In Fig. 6 we plot the spin polarization of the current versus the magnetic
field for different temperatures. The inset shows the spin-up and spin-down
components of MC. The spin-up components exhibit a maximum for small
magnetic field and decreases rapidly to zero, since the population for the
spin-up band decreases when the magnetic field increases. The maximum is due
to a maximum in the bandwidth (see Fig.~2(c)). From this inset we found that
the oscillation in the MC\ is due to the spin-down MC components. The spin
polarization at higher temperature increases more slowly than that at low
temperature due to the thermal fluctuations of the magnetization of the
magnetic ions.

\section{\bf CONCLUSIONS}

We studied the electron transport in DMS\ superlattices using a
semiclassical Boltzman equation, and investigated the effect of the s-d
exchange interaction which is treated using the molecular-field
approximation on the longitudinal spin transport in diluted magnetic
semiconductor superlattices. The conductivity exhibits an oscillating
behavior with varying magnetic field. The conductivity decreases rapidly for
small magnetic field, and increases for strong magnetic field. The dips in
the conductivity at strong magnetic fields are smeared out with increasing
temperature. The spin polarization increases rapidly with increasing
magnetic field and the longitudinal MC becomes spin-polarized in strong
magnetic fields. Our results clearly illustrate that one can adjust the
longitudinal spin transport by tuning the external magnetic field in DMS\
superlattices.

Most optical and transport properties of the band electrons in DMS were
successfully interpreted within the molecular-field approxiamtion. However,
one should keep in mind that this approach is only justified for pure
paramagnetic material where the every spin can be treated independently. In
paramagnetic DMS system the s-d exchange interaction induced spin splitting
is relevant for small barrier height, the spin-orbit interaction and
band-structure related spin-flip processes are not very efficient.\cite
{Bastard} Therefore in our calculation we ignored the effect of the
magnetization fluctuations which may be important at low magnetic fields
which is expected to increase the overall resistance and decrease the spin
polarization at low fields. These magnetic fluctuations are supressed in a
strong magnetic field (B%
\mbox{$>$}%
1T). Nevertheless, the previous investigations show that the theoretical
results based on the molecular-field approximation give the correct
positions for the maxima and minima of the MC in DMS\ systems\cite{ACC} in
strong magnetic fields since the magnetization fluctuations are supressed,
and therefore we believe that the qualitative behavior of our MC is correct
especially in strong magnetic fields (B%
\mbox{$>$}%
1T). In summary, the external magnetic field is a tool to tailor the
transport properties of DMS\ superlattices. Such systems are extremely
attractive from the point of view of both basic research and technological
application, such as for a spin filter.

\begin{acknowledgments}
This work is supported by the Chinese Academy of Sciences foundation,
the Chinese Science Foundation, the Flemish Science Foundation (FWO-Vl), 
the Interuniversity Microelectronics Center (IMEC), 
the Inter-University Attraction Poles (IUAP) research program,
the Flemish Concerted Action program (GOA),
and the Flemish-China bilateral science and technological
cooperation.
\end{acknowledgments}

\begin{figure}[tbp]
\caption{ The energy spectrum of the two lowest minibands of electron states
($n=0$) in a DMS superlattice for different spin orientations under two
different magnetic fields: a) B=0T and b) B=4T. The width of the barrier
(well) is $L_D=10$nm ($L_N=10$nm), the lattice period $L=L_D+L_N$ and $T=1K $%
. The solid curves denote the energy spectrum for the spin-up electron, the
dashed curves for the spin-down electron, the dotted line is the Fermi
energy, the thick solid and dashed curve denote the derivative of the Fermi
distribution and the Fermi distribution, $\partial f / \partial {\cal E}$
and $f({\cal E})$, near the Fermi energy, respectively. The band width of
the two lowest spin-up (Fig. 2(c)) and spin-down (Fig. 2(d)) bands are shown
in Figs. 2(c) and 2(d) as function of the magnetic field. The shaded regions
in the figures denote the electron miniband in the DMS superlattice. Fig.
2(e) shows the group velocity of the spin-up and spin-down electron lowest
miniband as a function of the momentum $k_z$ for different magnetic fields.}
\end{figure}

\begin{figure}[tbp]
\caption{The energy spectrum of the lowest miniband for different Landau
levels for spin-up and spin-down electrons as a function of magnetic field.
The thin solid lines represent the energy of spin-up electrons at $k_z=0$,
the thin dashed lines for spin-up electrons at $k_z=\pi/L$. The very thick
solid and dashed lines show the energies for the spin-down electron at $%
k_z=0 $ and $k_z=\pi/L$. The thickest solid curve shows the Fermi energy as
a function of magnetic field. $L_D=10$nm, $L_N=10$nm and the density of the
electrons $n_e=2\times10^{17}/cm^3$. }
\end{figure}

\begin{figure}[tbp]
\caption{ The conductivity $\sigma/\sigma_0 $ versus the magnetic field for
different temperatures, where $\sigma_0=n_{e}e^{2}\tau/m^{*}$. $L_D=10$nm, $%
L_N=10$nm and the density of the electrons $n_e=2\times10^{17}/cm^3$. The
inset shows the Fermi energy as a function of magnetic field.}
\end{figure}

\begin{figure}[tbp]
\caption{ The conductivity $\sigma/\sigma_0$ versus the magnetic field for
different densities. $L_D=10$nm, $L_N=10$nm, $T=1K$. The inset shows the
Fermi energy for two different densities.}
\end{figure}

\begin{figure}[tbp]
\caption{ The spin polarization of the current in DMS superlattice for
different temperatures. The inset shows the spin-up and spin-down MC
components as a function of magnetic field. The arrows in the inset
represent the spin-up and spin-down MC components. $L_D=10$nm, $L_N=10$nm, $%
T=1K$.}
\end{figure}


\begin{references}
\bibitem[{*}]{kai}  Electronic address: kchang@red.semi.ac.cn.

\bibitem[{*}*]{francois}  Electronic address: peeters@uia.ua.ac.be.

\bibitem{review}  {\it Diluted Magnetic Semiconductors}, edited by J. K.
Furdyna and J. Kossut, Semiconductors and Semimetals Vol. 25 (Academic, New
York, 1988).

\bibitem{Dai1}  N. Dai, H. Luo, F. C. Zhang, N. Sarmarth, M. Dobrowolska,
and J. K. Furdyna, Phys. Rev. Lett. {\bf 67}, 3824 (1991).

\bibitem{Dai}  N. Dai, L. R. Ram-Mohan, H. Luo, G. L. Yang, F. C. Zhang, M.
Dobrowolska, and J. K. Furdyna, Phys. Rev. B {\bf 50}, 18153 (1994)

\bibitem{Lunn}  M. Oestreich, J. Hubner, D. Hagele, P. J. Kar, W. Heimbrodt,
and W. W. Ruhle, D. E. Ashenford and B. Lunn, Appl. Phys. Lett. {\bf 74},
1251 (1999).

\bibitem{Kossut}  J. Kossut, Phys. Status Solidi (b) {\bf 72}, 359 (1975).

\bibitem{Furdyna}  J. K. Furdyna, J. Appl. Phys. {\bf 64}, R29 (1988).

\bibitem{von}  M. von Ortenberg, Phys. Rev. Lett. {\bf 49}, 1041 (1982).

\bibitem{SPT}  G. A. Prinz, Phys. Today {\bf 48}, 58 (1995); G. A. Prinz,
Science {\bf 282}, 1660 (1998).

\bibitem{Kikkawa}  J. M. Kikkawa, I. P. Smorchkova, N. Samarth, and D. D.
Awshalom, Science {\bf 277}, 1284 (1997); J. M. Kikkawa, and D. D. Awshalom,
Nature (London) {\bf 397}, 139(1999).

\bibitem{Fiederling}  R. Fiederling, M. Keim, G. Reuscher, W. Ossau, G.
Schmidt, A. Waag and Molenkamp, Nature {\bf 402}, 787 (1999).

\bibitem{Ohno}  Y. Ohno, D. K. Young, B. Benschoten, F. Matsukura, H. Ohno
and D. D. Awschalom, Nature {\bf 402}, 790 (1999).

\bibitem{Lend}  J. Hajdu and G. Landwehr, in {\it Strong and Ultrastrong
Magnetic Fields and Their Applications}, edited by F. Herlach
(Springer-Verlag, New York, 1985), pp. 17-112.

\bibitem{Poly}  V. M. Polyanovskii, Sov. Phys. Semicond. {\bf 18}, 1142
(1984).

\bibitem{Datar}  A. E. Datars and J. E. Sipe, Phys. Rev. B {\bf 51}, 4312
(1995).

\bibitem{Bastard}  G. Bastard and R. Ferreira, Surf. Sci. {\bf 267}, 335
(1992); G. Bastard and L. L. Chang Phys. Rev. B {\bf 41} 7899 (1990).

\bibitem{ACC}  A. Lemaitre, C. Testelin, C. Rigaux, T. Wojtowicz and G.
Karczewski, Phys. Rev. B {\bf 62}, 5059 (2000).

\bibitem{Rossin}  V. V. Rossin, F. Henneberger, and J. Pulsm, Phys. Rev. B 
{\bf 53}, 16444 (1996).

\bibitem{SLee}  S. Lee, M. Dobrowolska, J. K. Furdyna, H. Luo and L. R.
Ram-Mahan, Phys. Rev. B {\bf 54}, 16939 (1996); S. Lee, M. Dobrowolska, J.
K. Furdyna, and L. R. Ram-Mohan, Phys. Rev. B {\bf 59}, 10302 (1999).

\bibitem{Egues}  J. C. Egues, Phys. Rev. Lett. {\bf 80}, 4578 (1998).

\bibitem{Burt}  The formulism used in our calculation can be found in the
work of F. Szmulowicz [Phys. Rev. B {\bf 54}, 11539 (1996)] which is based
on the exact envelope-function formulism developed by M. G. Burt [J. Phys.
Condens. Matter {\bf 4}, 6651 (1992). In the case of a one-band model, it
reduced to the familiar Kronig-Penney model result (G. Bastard, Phys. Rev. B 
{\bf 25}, 7584 (1982)).

\bibitem{PALee}  P. A. Lee and T. V. Ramakrishnan, Rev. Mod. Phys. 57, 287
(1985).

\begin{figure}[tbp]
\caption{Schematic illustration of a ZnSe/Zn$_{0.96}$Mn$_{0.04}$Se DMS
superlattice subjected to a perpenticular magnetic field. The shaded regions
denote the diluted magnetic semiconductor layers. Fig. 1 (a) shows the
potential profile for an electron in a DMS superlattice in the absence of a
magnetic field, Figs. 1 (b) and (c) show the potential profiles for the
spin-up and spin-down electron in the presence of a magnetic field,
respectively. The probabilities for the spin-up and spin-down electrons are
plotted in the figure. }
\end{figure}
\end{references}
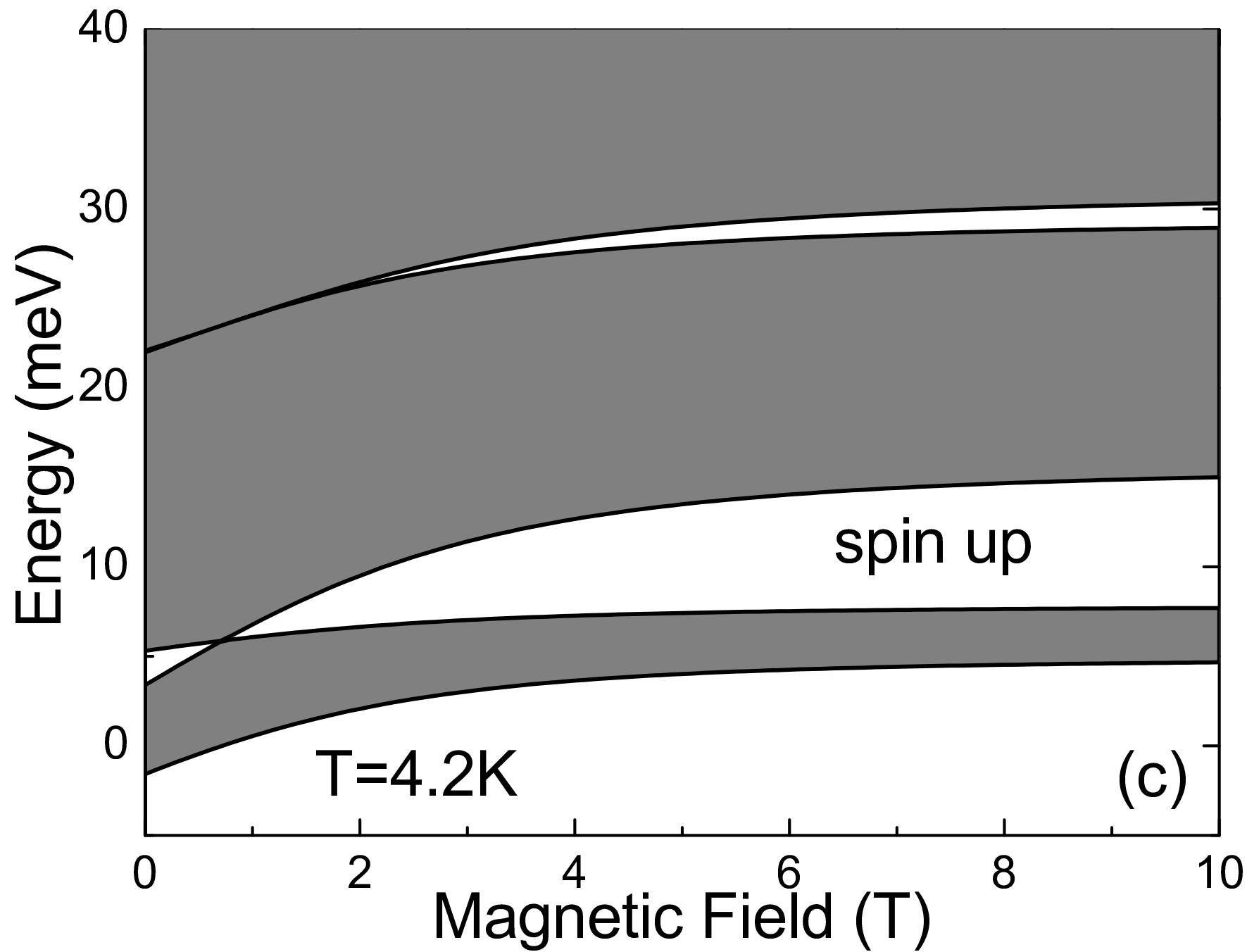
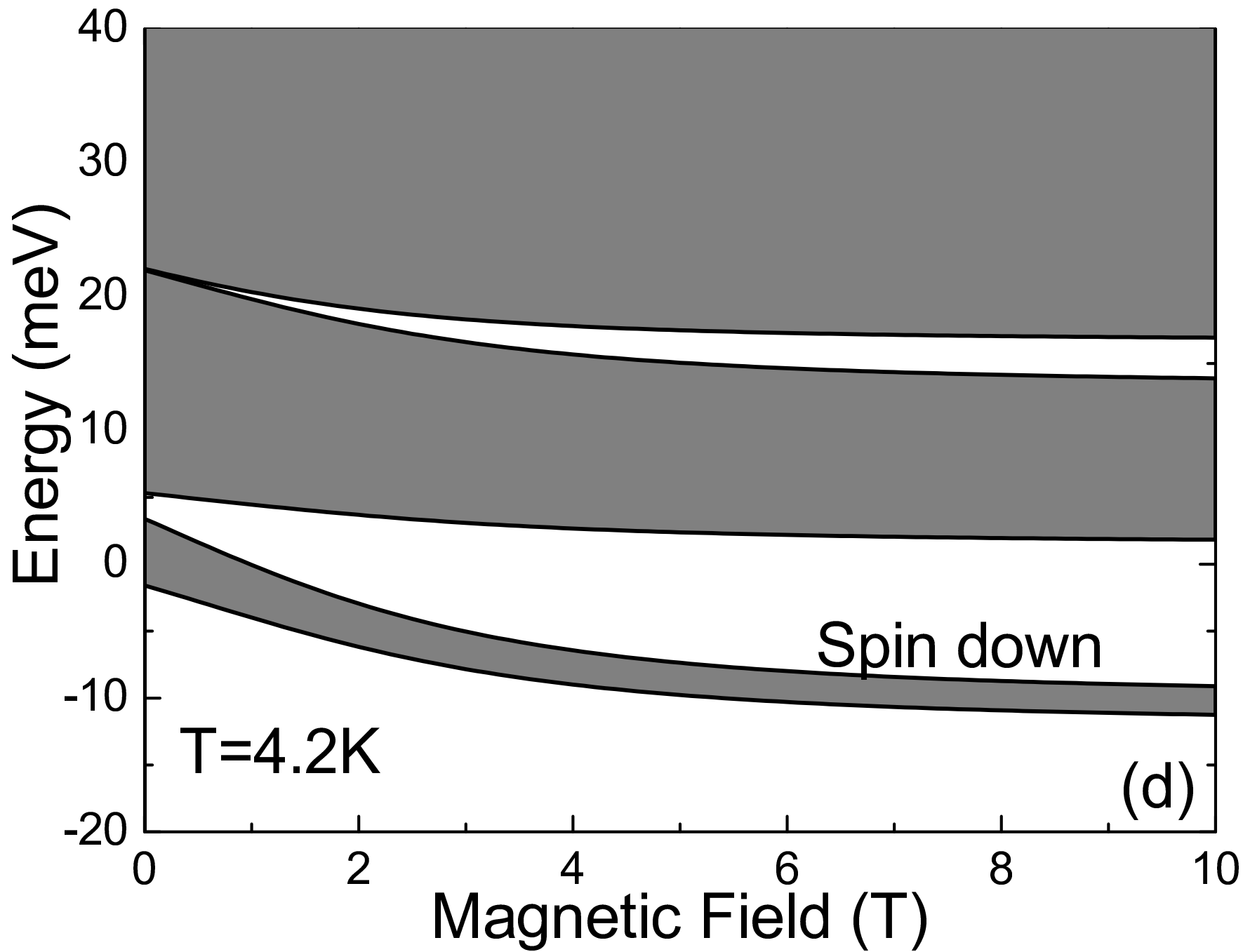
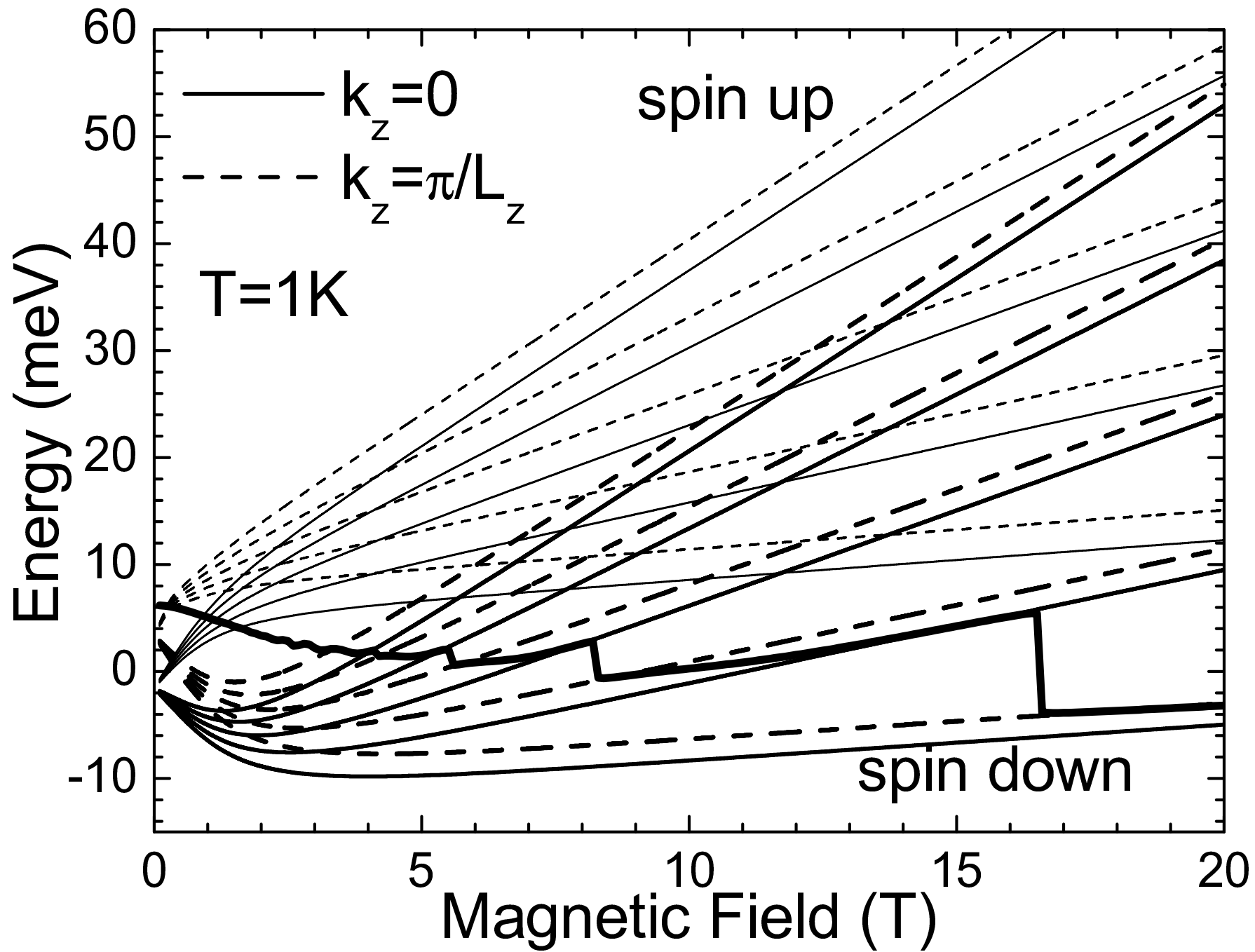
\end{document}